\documentstyle[amssymb,preprint,prl,aps,12pt]{revtex}
\begin{document}
\title{{\bf Slow relaxation in weakly open vertex-splitting rational polygons}}
\author{Valery B. Kokshenev$^{1}${\large \ }and Eduardo Vicentini$^{2}$}
\address{$^{1}${\it Departamento de F\'{i}sica, Universidade Federal de Minas Gerais,}%
\\
ICEx, Caixa Postal 702, CEP 30123-970, Belo Horizonte, MG, Brazil\\
$^{2}${\it Departamento de F\'{i}sica, Universidade Estadual do Centro}\\
Oeste, Caixa Postal 730, CEP 85010-990, Guarapuara, PR, Brazil\bigskip }
\date{\today }
\maketitle

\begin{abstract}
\leftskip 54.8pt \rightskip 54.8pt

The problem of splitting effects by vertex angles is discussed for
nonintegrable rational polygonal billiards. A statistical analysis of the
decay dynamics in weakly open polygons is given through the orbit survival
probability. Two distinct channels for the late-time relaxation of type $%
t^{\delta }$ are established. The primary channel, associated with the
universal relaxation of ''regular'' orbits, with $\delta =1$, is common for
both the closed and open, chaotic and nonchaotic billiards. The secondary
relaxation channel, with $\delta >1$, is originated from ''irregular''
orbits and is due to the rationality of vertices.


\end{abstract}


\section{INTRODUCTION}

Polygonal classical billiards is an active subject of research in
mathematics and physics (see for review Ref.\cite{G96}). In view of the null
Lyapunov exponent and the null Kolmogorov metric entropy the rational
polygons, formed by the piece-line billiard boundary with the vertex angles
are rational multiplies of $\pi $, are known to be {\em nonchaotic }systems%
\cite{G96,ZK75,RB81,EFV84,G86,M00}. They are therefore well distinct from
the Sinai billiard\cite{Sin79} (SB) and the Bunimovich billiard\cite{Bun79}
(BB) where classical {\em chaotic} motion regimes are due to, respectively, 
{\em dispersive\ effects} caused by the circle disk and the squared
boundary, and the {\em interplay\ }between boundary segments formed by the
circle and the square. Meanwhile, the rational polygons of $m$ equal sides
and equal vertices (hereafter, the $m$-gons\cite{G96}) revealed\cite{VUF93}
positive Lyapunov exponents with increasing of $m$. Furthermore, the
polygonal billiards exposed chaoticlike changes in the associated
quantum-level spectra\cite{CC89}, which fluctuations are shown\cite
{SS93,SS95} to be very close to the Gaussian orthogonal-ensemble-type
statistics. In view of the {\em splitting effects} by the angle vertices,
polygons do not satisfy the conditions of integrability\cite{ZK75,andK}

These evidences for the chaoticlike features of the nonintegrable rational
polygons were recently questioned by Mantica\cite{M00} through the orbital
complexity analysis. Unlike the case of the integrable billiards\cite{VK01},
a delicate problem of the interplay between the regular (piece-line) and the
irregular (vertex-angle) boundary segments in polygons cannot be solved in
terms of the first-order averaged polygonal characteristics, such as the
average-orbit coding length\cite{M00}, the mean wall-collision time\cite
{VK01}, or the average collision number\cite{KV02}. As shown through
analysis of the orbit-wall-collision statistics\cite{KV02}, the higher order
correlation effects induced by vertices play a crucial role in the intrinsic
dynamics of $m$-gons. Earlier this was experimentally corroborated\cite{SS93}
through the quantum-level spectra statistics: ''unlike the case of the
irrational polygons, the long-range level correlation effects are due to the 
{\em rationality} of vertex angles''\cite{SS93}. In the case of the
curved-by-circle infinite-$m$ rational polygon ($\infty $-gon) the late-time
memory effects arise from the {\em sliding orbits}\cite{KV02}, which have no
analog in the ballistic-type dynamics of its counterpart given by the circle
billiard (hereafter CB), where the orbit classification is well established%
\cite{Rob99}. This implies that the quasi-classical approach has no
justification for the ''quantized'' vertex-spliting $\infty $-gon that is
geometrically equivalent to the CB. In other words, the quantum-to-classical 
{\em dynamics correspondence} suggested\cite{M00} between a given $m$-gon
and its circumscribing counterpart is violated\cite{KV02} regardless of the
fact that the {\em geometric} {\em correspondence} exists and can be
achieved with any precision when $m\rightarrow \infty $ (with the help of
the aforementioned first-order averaged characteristics). This is in
agreement with a conclusion on inapplicability in polygons of the
quantum-to-classic correspondence principle. The latter was elaborated\cite
{SS95} within the scope of the conventional Wentsell-Kramer-Brillouin
picture that failed to establish a one-to-one correspondence between
classical orbits and their quantum counterparts.

The issue of the current paper is an investigation of the vertex-splitting
effects revealed in the {\em \ decay} dynamics of finite-$m$ rational
polygons. An analysis is given through the survival orbit statistics in the
weakly open billiards, which boundaries permit orbits to escape through a
small opening. We will see that, similarly to the case of the intrinsic
dynamics in the closed $m$-gons\cite{KV02}, the vertex-splitting effects are
dual of vertex-ordering and vertex-disordering motion effects, that is
manifested through, respectively, the orderlike and chaoticlike behavior
observed in the late-time relaxation. The paper is organized as follows. The
recent findings for the decay dynamics in nonchaotic\cite{VK01} and chaotic%
\cite{KN00} billiards are given in Sec.II and discussed within concepts of
the primary and the secondary relaxation channels. The weakly open rational
polygons are analyzed numerically and analytically in Sec.III for the cases
of small and large numbers $m$. Discussion and conclusions are summarized in
Sec.IV.

\section{DECAY DYNAMICS OF THE CHAOTIC AND NONCHAOTIC BILLIARDS}

The intrinsic dynamics of the closed classical billiards is commonly
considered in terms of a temporal decay of the correlation functions for
certain dynamical variables (see e.g. Ref.\cite{GG94}). Pure exponential
lost of amount of memory on the initial states is not a unique channel of
relaxation even in the chaotic systems (see e.g. Ref.\cite{ACG96}). By
studying the chaotic billiards, such as the SB\cite{ACG96,MZ83,FM84,ZGN86},
that is dynamically equivalent to the correspondent Lorentz gas (LG) model%
\cite{GG94} and the BB\cite{Bun79,Alt96,VCG83} with a stadium geometry, it
has been recognized that a crossover from the short-time exponential to the
late-time algebraic decay is due to the long-term memory effects on a free
regular motion. The algebraic tail of the correlation functions seems to be
vanished only in the case of the fully hyperbolic systems correspondent to
such geometries as the finite-horizon SB\cite{GG94,FM84} (equivalent to the
high-density LG) or the diamond\cite{GG94}. Qualitatively the same can be
referred to the {\em \ decay dynamics} in weakly open billiards that
describes a crossover from a bounded to unbounded orbit free motion. Such a
decay dynamics is initially established by uniformly distributed $N_{0}$
point particles (of unit mass and unit velocity) moving inside of the closed
planar billiard table and then allowing to escape through a small opening of
width $\Delta $. A temporal behavior of the dynamic observables can be
scaled to the characteristic billiard times, namely 
\begin{equation}
\tau _{c}=\frac{\pi A}{P}\text{ and }\tau _{e}=\frac{P}{\Delta }\tau _{c}%
\text{.}  \label{tau/ec}
\end{equation}
Here the {\em mean collision time}\cite{GG94,Che97,Gar97} $\tau _{c}$ and by
the {\em mean escape time}\cite{VK01,GG94,Alt96,BB90} $\tau _{e}$ are
introduced through the accessible area $A$ and the perimeter $P$ ($\gg
\Delta $) for a given billiard table. The late-time ($t\gg \tau _{e}$)
algebraic-type evolution of non-escaped orbits (or particles) $%
N_{t}\varpropto N_{0}(\tau _{e}/t)^{\delta }$, which time derivation is
related to the {\em survival probability}, is definitely characterized by
the {\em decay} {\em dynamic exponent} $\delta $.

In the integrable {\em nonchaotic }billiards, the first observation of the
algebraic decay with exponents $\delta \lessapprox 1$ was given\cite{BB90}
for the case of the square billiard. Recent study of the decay dynamics in
the $4$--gon revealed\cite{VK01} two distinct channels of the algebraic-type
slow relaxation. The first one is due to the regular-orbit motion with the
decay dynamic exponent $\delta =1$, and the second channel is originated
from the ''irregular'' orbits induced by the singular vertex-spliting
effects, which give rise to the subdiffusion motion regime indicated\cite
{VK01} by $\delta \thickapprox 0.85$. Such a kind of diverse dynamical
behavior was observed through the {\em \ survival orbit spectra} defined by
a number for the survived orbits and simulated in the integrable CB and in
the almost-integrable\cite{G96} square billiard (see, respectively, Figs.
5,6 and 3,4 in Ref.\cite{VK01}). In both the cases an irregularlike motion
is due to orbit families known as, respectively, the ''whispering-gallery''
(polygonlike) and the ''bouncing-ball'' orbits. Meanwhile, in the late-time
decay of the CB (see inserts in Figs. 3,4 and 5,6 in Ref.\cite{VK01}) the
short-time living, frequently escaped ''whispering-gallery'' orbits, unlike
the nonintegrable case, did not contribute to the second channel of
relaxation. Thus, the solely exponent $\delta =1$ was observed in the CB. 

In the {\em chaotic} closed and weakly open (including Hamiltonian)
classical systems, presented by the BB\cite{Alt96,VCG83,Bun85,Pik92}, the
infinite-horizon SB\cite{KN00,ACG96,MZ83,ZGN86,FRS94,FS95}, and by the
low-density LG model\cite{FM84,Ble92,Dah96}, the algebraic-type decay was
numerically revealed\cite{FKC01} by the dynamic exponents $\delta \geq 1$.
Similarly to the nonchaotic case, it has been repeatedly recognized that the
algebraic tail is caused by the ''arbitrary long segments'' observed in the
evolution of stochastic orbits\cite{VCG83}, or by the regularlike orbit
motion due to ''sticking particles'' \cite{Pik92,HBS92}. This implies that
in both the cases this relaxation is due to a free motion of the
corresponding trajectories in the infinite (with respect to the observation
time) distinct{\em \ corridors} which are open in the relevant phase space%
\cite{FRS94,FS95,Ble92}.

The algebraic-type relaxation channel with $\delta =1$ ($\equiv \alpha $)
established in the chaotic\cite{KN00,ACG96,Alt96,FRS94,FS95} as well as in
nonchaotic\cite{BB90,VK01} billiards, seems to be generic for all non-fully
hyperbolic systems with smooth convex boundaries. Its independence of the
billiard space dimension\cite{FS95}, its insensitiveness to details of the
boundary shape\cite{VK01}, including that to a position of the small opening%
\cite{Alt96}, and to the initial conditions\cite{VK01}, suggests that the
late-time $\alpha $-relaxation arise in classical systems as the {\em %
universal primary relaxation}. The latter is a part of the two-step
relaxation scenario discussed in the chaotic\cite{KN00} and nonchaotic\cite
{VK01} weekly open classical systems. This {\em universal scenario} was
introduced by the short-time pure exponential and by the
first-power-algebraic overall orbit decays given by $N_{t}\varpropto
N_{0}e^{-(t/\tau _{e})^{\gamma }}$ with $\gamma =1$, and by $N_{t}\varpropto
N_{0}(\tau _{e}/t)^{\delta }$ with $\delta =1$, respectively. Meanwhile the
real relaxation does not exclude the existence of the additional
intermediate {\em transient} {\em regimes} approximated by the
stretched-exponential form\cite{KN00} with $\gamma <1$ (earlier discussed
for chaotic billiards in analytical\cite{BS81} and numerical\cite{GG94}
forms), and by other algebraic-decay forms\cite{FRS94,FS95} with $\delta
\lessgtr 1$. The escape mechanism of the primary relaxation in the chaotic
and nonchaotic billiards was described in details within a coarse-grained
approximation (see, respectively, Eq.(12) in Ref.\cite{KN00} and Eq.(15) in
Ref.\cite{VK01}). Unlike the case of the primary relaxation, the temporal
observation conditions (observation windows) for the {\em secondary
relaxation }in the chaotic billiards ($\delta =\beta >1$) are shown to be
very sensitive to the billiard geometry\cite{VK01}, to the dimension\cite
{FS95} $d$ of a billiard table, and to the initial conditions\cite{VK01}.
This can be exemplified by the dynamic-exponent constraint $1<\beta \leq d$
proposed in Ref.\cite{FS95} and observed in the chaotic BB\cite{Alt96} and SB%
\cite{ACG96,FRS94,FS95} billiards.

The algebraic-type decay of correlations in the low-density {\em chaotic} LG
is due to evolution of trajectories within the infinite principal and/or
''hindered'' open corridors\cite{FRS94,FS95,Ble92,Dah96}. In the
corresponding disk-dispersing SB (of side $L$ and disk radius $R$) with the
infinite-horizon geometry ($R<L/2$) the principal and the ''hindered''
corridors, respectively, can be governed by the limiting disk radii\cite
{Ble92} $R_{\alpha }=L/2$ and $R_{\beta }=\sqrt{2}L/4$ and observed through
the established\cite{KN00} algebraic-tail temporal windows. Thus, the
primary relaxation was observed\cite{Diss} within the domain $R_{\beta }$ $%
<R<$ $R_{\alpha }$, when the ''hindered'' corridors are closed\cite{under}.
Under the geometric conditions $0<R$ $\leq R_{\beta }$ the transient $\beta $%
-relaxation channel was activated and indicated by the decay dynamic
exponents $\beta =1.2$ and $\beta =1.1$, respectively, in Refs.\cite{FRS94}
and \cite{Diss}. One can see that the primary and the secondary relaxation
channels can be geometrically associated with Bleher's\cite{Ble92} principal
and ''hindered'' corridors, respectively. The latter case was additionally
characterized\cite{Ble92} as a {\em superdiffusive} motion regime, observed
earlier in the chaotic\cite{GG94,MZ83,ZGN86,Dah96} and very recently in the
polygonal\cite{KV02} closed billiards.

We see that non-fully hyperbolic billiards, on one hand, are indiscernible
within decay dynamics observed through the primary universal $\alpha $%
-relaxation channel. On the other hand, the chaotic and nonchaotic, the open
and closed billiards are well distinguished with respect to the secondary $%
\beta $-relaxation given by the decay dynamic exponents $\beta >1$ and $%
\beta <1$, respectively. In what follows we give numerical and analytical
analysis of the conditions of stabilization of both the primary and
secondary relaxation channels in the weakly open rational polygonal
billiards.

\section{ORBIT DECAY IN POLYGONS}

\smallskip We deal with the rational polygons{\em \ }of $m$ equal sides,
denominated as $m$-gons, circumscribed below a circle of radius $R$. The
mean collision time $\tau _{cm}=(\pi R/2)\cos (\pi /m)$ and the mean escape
time $\tau _{em}=(\pi R^{2}m/2\Delta )\sin (2\pi /m)$ are given with the
help of Eq.(\ref{tau/ec}) through area $A_{m}=(mR^{2}/2)\sin (2\pi /m)$ and
perimeter $P_{m}=2mR\sin (\pi /m)$. In the limit $m\rightarrow \infty $ one
naturally arrives at the circle geometry of the $\infty $-gon with the mean
times $\tau _{c\infty }=\tau _{cR}^{(CB)}=\pi R/2$ and $\tau _{e\infty
}=\tau _{eR}^{(CB)}=\pi ^{2}R^{2}/\Delta $ , both characteristic of the CB.
This demonstrates how the first-order dynamic characteristics can be
introduced through the aforementioned geometrical correspondence that takes
place between the $\infty $-gon and the CB. Meanwhile the dynamical
correspondence does not exist\cite{KV02} because the vertex-memory effects
violate commutation between of the temporal ($t\rightarrow \infty $) and
spatial ($m\rightarrow \infty $) limits.

\subsection{Small Number of Vertices}

Similarly to the closed $m$-gons\cite{KV02}, let us consider the case of
small number of vertices, with $m<10$ , within the scope of the
deterministic approach. This is given by generalization of the regular-orbit
description introduced\cite{VK01} for the particular case of $m=4$ and is
straightforwardly based on the fact that the wall-collision angles $\varphi $
(counted off the normal to the boundary and preserved by elastic refections)
are integrals of motion. This is true for the integrable billiards where
dispersing or splitting effects are absent. A dynamic description of the
regular-orbit motion can be introduced with accounting of the fact that $m$
(or $m/2$) sides of a given $m$-gon, with odd (or even) number of vertices,
are dynamically equivalent\cite{DynEqui}. A description of the
wall-collision statistics can be therefore reduced to the collision-angle
domain $\varphi =[0,\varphi _{m}]$ with 
\begin{equation}
\varphi _{m}\text{ {}}=\left\{ 
\begin{array}{cc}
\pi /2m & \text{ for odd }m\text{{},} \\ 
\pi /m & \text{for even }m.\text{{}}
\end{array}
\right.   \label{fi-m}
\end{equation}
In turn, the $\varphi $-family regular--orbit sets can be introduced (for
details see Appendix) through the characteristic collision times, namely 
\begin{eqnarray}
t_{cm}(\varphi ) &=&\frac{\pi R}{2\varphi _{m}}\frac{\cos \pi /m\text{ }\sin
\varphi _{m}}{\cos (\varphi -\psi _{m})}\text{ , with}  \label{tc-m} \\
\psi _{m} &=&\left\{ 
\begin{array}{cc}
0 & \text{for odd }m\text{{} and }m/2, \\ 
\pi /m & \text{for even }m\text{{}}/2.
\end{array}
\right.   \label{psi-m}
\end{eqnarray}
The collision time $t_{cm}(\varphi )$ is related to the billiard mean
collision time $\tau _{cm}$  through the ''mean-collision-time'' equation, 
{\it i.e.}, $<t_{cm}(\varphi )>_{c}\equiv \int_{0}^{\varphi
_{m}}t_{cm}(\varphi )f_{0m}(\varphi )d\varphi =$ $\tau _{cm}$. The latter is
equivalent to the known\cite{Che97} ''mean-free-path'' equation considered
in the uniformly populated two-dimensional ($2D$){\em \ collision }subspace $%
\Omega _{cm}$ (see also Eqs.(3,4) in Ref.\cite{VK01}) the $3D$ phase space $%
\Omega _{m}$. The $\varphi $-set-orbit distribution function $f_{0m}$ is
defined here as 
\begin{equation}
f_{0m}(\varphi )=\left\{ 
\begin{array}{cc}
\frac{\cos (\varphi -\psi _{m})}{\sin \varphi _{m}}\text{, } & \text{for
subspace }\Omega _{cm}, \\ 
\frac{t_{cm}(\varphi )}{\tau _{cm}}\frac{\cos (\varphi -\psi _{m})}{\sin
\varphi _{m}}\equiv \frac{1}{\varphi _{m}}\text{,} & \text{for space }\Omega
_{m}
\end{array}
\right.   \label{f-spaces}
\end{equation}
by generalization of Eqs.(6,7) in Ref.\cite{VK01}.

We discuss the late-time ($t\gg \tau _{em}$) survival dynamics in a given $m$%
-gon through the $\varphi $-set-orbit decay spectra defined by numbers of
the survived orbits $N_{m}(t,\varphi )$ and by corresponding overall numbers 
$N_{tm}=<N_{m}(t,\varphi )>_{c}$. Here a procedure of averaging over
collisions, denoted by $<...>_{c}$, is introduced through the aforesaid
''mean-collision-time'' equation. The universal relaxation channel,
associated solely with the regular orbits, is given through numbers of the
nonescaped orbits, namely 
\begin{equation}
\text{ }\frac{N_{m}(t,\varphi )}{N_{0m}}=C_{m}(\varphi )\frac{t_{cm}(\varphi
)}{\tau _{cm}}f_{0m}(\varphi )\frac{\tau _{em}}{t}\text{ and }\frac{N_{tm}}{%
N_{0m}}=D_{m}\frac{\tau _{em}}{t}\text{,}  \label{Nt}
\end{equation}
which are the late-time solutions of the relevant decay-kinetics master
equation (see Eqs. (16,17) in Ref.\cite{VK01}). As seen from Eq.(\ref{Nt}),
the fundamental characteristics $t_{cm}(\varphi ),$ $f_{0m}(\varphi )$ and $%
\tau _{cm}$ are common for both the decay and intrinsic dynamics. The $%
\varphi $-set orbit-partial weight $C_{m}(\varphi )$ and the orbit-overall
weight $D_{m}$ of the corresponding algebraic tails can be established in
explicit form within a certain coarse-grained scheme and directly observed%
\cite{VK01,KN00}. However an interesting analysis that indicates a departure
of $m$-gons from the true integrable systems due to vertices can be given
through the algebraic-tail weights without detail calculations.

By taking into account the mentioned equation $N_{tm}=<N_{m}(t,\varphi )>_{c}
$, the overall regular-orbit-decay weight $D_{cm}=C_{cm}^{(\text{reg}%
)}=<C_{m}(\varphi )>_{c}^{(\text{reg})}$ immediately follows from Eq.(\ref
{Nt}). The validity of this relation can be straightforwardly tested for the
integrable CB where the irregular orbits do not survive in the late-time
relaxation. Indeed, in this case $C_{cR}^{(\exp )}=0.206$ and $D_{cR}^{(\exp
)}=0.214$ that follows from the observation (see Tab.2 in Ref.\cite{VK01})
of the primary relaxation of the collision space $\Omega _{cR\text{ }}$. One
can see that the algebraic-tail parametrization of the regular-orbit decay
effects in the CB is self-consistent that is experimentally justified with a
high precision, {\it i.e.}, $C_{cR}^{(\exp )}=D_{cR}^{(\exp )}=0.210\pm 2\%$%
. In the case of polygons, a violation of the relations $C_{cm}^{(\text{reg}%
)}=C_{cm}^{(\exp )}$ and $C_{cm}^{(\exp )}=D_{cm}^{(\exp )}$ is expected in
view of the long-living irregular-orbit motion induced by vertex-splitting
effects.

We have performed numerical experiments\cite{PCexp} on decay dynamics in $m$%
-gons with small number of vertices: $m=3,4...8$. The initial particles ($%
N_{0m}=10^{6}$) have been distributed randomly within the two distinct phase
spaces described in Eq.(\ref{f-spaces}) and then allowed to escape through a
small opening $\Delta \ll R$. The condition $\tau _{em}$ ($=300$) is chosen
to be common for all $m$ and that has been provided with the accuracy of $%
\pm 5\%$ with the help of Eq.(\ref{tau/ec}). In all cases of $m\leq 8$ the
typical algebraic decay is observed\cite{Diss} within a certain temporal
windows given by, approximately, $10^{1}\tau _{em}<t\leq 10^{3}\tau _{em}$.
The particular cases of the observed decay spectra for pentagon and heptagon
are exemplified in {\bf Fig.1}. In general, the overall-orbit late-time
decay in $m$-gons with small number of vertices do not show noticeable
deviations from the linear relaxation\cite{exept} (see the left insert in
Fig.1). Thus the partial weights $C_{m}^{(\exp )}(\varphi )$ are derived
from the observed numbers $N_{m}^{(\exp )}(t,\varphi )$ through Eq.(\ref{Nt}%
) with accounting of the estimated distribution function $f_{0m}(\varphi )$
and the collision times $t_{cm}(\varphi )$ given in, respectively, Eqs.(\ref
{fi-m}) and (\ref{tc-m}). These equations were additionally experimentally
tested\cite{Diss} ({\it e.g.} see the right insert in Fig.1).

As seen from Fig.1, the observed partial weights $C_{m}^{(\exp )}(\varphi )$
exhibit regular (small) and irregular (large) deviations from the mean
magnitude $C_{cm}^{(\exp )}=C_{cm}^{(\text{tot})}$ shown by a solid
(horizontal) line. The latter and the regular-orbit weights $C_{cm}^{(\text{%
reg})}$ (evaluated with regardless of the large isolated peaks) are
accumulated in {\bf Table.1}. Through analysis of a difference between the
overall and partial weights $\Delta C_{cm}=C_{cm}^{(\text{tot})}-C_{cm}^{(%
\text{reg})}$, one can see with the help of Table 1 that, similarly to the
case of the intrinsic dynamics (see Fig.2 in Ref.\cite{KV02}), the
vertex-splitting effects in the even-gons are more pronounced than those in
the odd-gons. In all the $m$-gon cases a deviation of the total weights,
treated as $\Delta D_{cm}=D_{cm}^{(tot)}-D_{cm}^{(\text{reg})}$ (with $%
D_{cm}^{(\text{tot})}=D_{cm}^{(\exp )}$ and $D_{cm}^{(\text{reg})}=C_{cm}^{(%
\text{reg})}$ given in Table 1), exceeds experimental error ($\pm 2\%$)
established above for the integrable SB, and we therefore infer that $%
D_{cm}^{(tot)}>D_{cm}^{(\text{reg})}$. This implies that the irregular-orbit
motion is involved into the observed relaxation. On the other hand, no
noticeable deviation from the primary relaxation is indicated in the
observed decay-orbit dynamics when $m\leq 8$. Similarly to the case of the
intrinsic dynamics observed\cite{M00,KV02} in the $m$-gons with small number
of vertices, we deduce that the regular-orbit motion dominates in the
late-time decay dynamics.

\subsection{Large Number of Vertices}

The universal two-step relaxation in open billiards is shared by $m$-gons
with arbitrary number of vertices\cite{Diss}. In {\bf Figs. 2}{\large \ }and 
{\bf 3 }we analyze our numerical results for the late-time overall-orbit
decay dynamics in the $m$-gons with large number ($m=2^{n}$ with $n=3,...6$)
of vertices for the cases of relatively small and large opening widths. In
general, one can observe that the decay dynamics of the rational polygons
with increasing of number of sides moves away from that given by the
geometrically corresponding CB: the chaotic effects, manifested by the
secondary relaxation channel with $\beta >1$, become more pronounced with
number of vertices. As seen from the observed relaxation of the initially
equivalent states (given in phase spaces $\Omega _{\infty }$ and $\Omega _{R}
$) is qualitatively distinct, and the {\em open} $\infty $-gon and the CB
are not therefore dynamically equivalent. Moreover, the $m$-gons with $m>8$
do not expose the algebraic decay with $\beta <1$, characteristic for {\em %
subdiffusive} motion regimes in nonchaotic open systems observed\cite{VK01}
in $4$-gon. 

In the particular case of $\Delta =0.05R$ shown in Fig. , the universal
relaxation remains stable until $m=64$, but when $m\geq m_{\alpha }^{(\exp
)}=128$ the primary relaxation channel turns up to be closed. In other
words, the regular-orbit-motion relaxation affected by vertices is
transformed into the irregularlike-motion relaxation indicated by dynamic
exponent $\beta =1.2$. Qualitatively the same follows from Fig. 3, but the
upper limit for the $\alpha $-channel observation window shows its
dependence on $\Delta $, {\it i.e.,} $m_{\alpha }^{(\exp )}=32$ for $\Delta
=0.20R$. \smallskip Thus Bleher's principal corridors of the regular-motion
relaxations are open for $3\leq m<m_{\alpha }^{(\exp )}$ (similarly to the
case of the SB given by $R_{\beta }$ $\leq R<$ $R_{\alpha }$ that is
discussed in Sec.II) when the vertex-splitting (or disk-dispersing effects)
are weak\cite{KV02}. We see that the $\alpha $-relaxation ( $3\leq
m<m_{\alpha }^{(\exp )}$) occurs as a precursor of the $\beta $-relaxation
regime, realized for $m\geq m_{\alpha }^{(\exp )}$, just as in the chaotic
SB case the $\beta $-relaxation ($0<R$ $\leq R_{\beta }$) was observed\cite
{Diss} before the $\alpha $-relaxation ($R_{\beta }$ $\leq R<$ $R_{\alpha }$%
). Conversely, the observed $\alpha $-to-$\beta $-relaxation crossover in
the $m$-gons, induced by the interplay between the piece-line regular and
the vertex-angle singular boundaries, is similar to that between the
semi-square and semi-circle parts of the stadium boundary. The latter was
deduced\cite{Alt96} in the case of the BB with small ($\Delta =0.01$) and
large ($\Delta =0.25$) opening widths, with the observed exponents $%
1\lessapprox \beta \lessapprox 2$ (see Table 1 in Ref.\cite{Alt96}).

Qualitatively, the effect of closing of the principal corridors in a given $m
$-gon can be understood by difficulties, increasing with $m$, to draw the
long segments of free motion, which intersect the polygonal sides in the
correspondent LG lattice but avoid the vertex angles (see also discussion in
Appendix). By contrast, the  $\beta $-relaxation revealed in Figs. 2, 3 is
associated with stabilization of the ''irregular''-type trajectories, which
are effectively deviated by vertex angles. More precisely, the observed
order-to-chaoticlike crossover can be understood as a regular-to-irregular
orbit transformation of the aforesaid sliding orbits (formed\cite{KV02} by $%
\varphi \thickapprox \pi /2$ -sets with characteristic free-path times $%
t_{cm}^{(reg)}(\varphi )=\tau _{c\infty }\cos ^{-1}\varphi $ following from
Eq.(\ref{tc-m})), which survive in the open $m$-gons with $m<m_{\alpha }$,
into those, renormalized substantially by vertex angles, called\cite{KV02}
by {\em vortexlike} orbits (with the large, but finite mean characteristic
time $\tau _{cm}^{(irreg)}=\tau _{c\infty }m/\pi $), which expected to be
stable for $m\geq m_{\alpha }\gg 1$. Within this context, the survival
conditions of the regular (sliding) and the irregular (vortexlike) orbits
driven, respectively, by piece-line and vertex-angle parts of the open
polygonal boundary, can be introduced as follows. On one hand, the favorable
observation conditions for the $\alpha $-relaxation (or the $\beta $%
-relaxation) should ensure to exclude (or to include) the vertex-angle
effects under the constraint $m<m_{\alpha }$ (or $m\geq m_{\alpha }$). On
the other hand, in the weakly open ($\Delta \ll P_{m}$) $m$-gon of a side
length $L_{m}=P_{m}/m$, the survival conditions for the regular sliding or
irregular vortexlike orbits are satisfied by geometric constraints,
respectively, $\Delta \ll L_{m}$ or $\Delta \gg L_{m}$. Hence, the $\alpha $-%
$\beta $-relaxation crossover, observed at $m=m_{\alpha }$, is ensured by
the condition $\Delta =L_{m}$. With taking into account that perimeter in
the polygons with large number of sides is well approximated by $%
P_{m}\thickapprox 2\pi R$, one arrives at the desirable criterium 
\begin{equation}
m_{\alpha }=\frac{2\pi R}{\Delta }.  \label{m-alfa}
\end{equation}
This finding provides the estimates $m_{\alpha }^{(\text{theor})}=126$ and $%
31$ for the particular cases of $m_{\alpha }^{(\exp )}=128$ and $32$
realized in Figs. 2 and 3, respectively. We infer that unlike the case of
the closed polygons\cite{KV02}, the vertex-splitting effects in the open $m$%
-gons with large number of vertices give rise to stabilization of the
vortexlike-orbit motion.

\section{DISCUSSION AND CONCLUSIONS}

The mild discontinues caused by vertex angles and relative lengths of the
edges is the central problem of the intrinsic dynamics of the ''almost
integrable'' polygonal billiards commonly discussed\cite{G96} in terms of
the orbit ergodicity, mixing, entropy, coding, complexity\cite{M00},
pseudo-integrability\cite{RB81}, spectral level statistics\cite{CC89,SS93},
and the orbit collision statistics\cite{KV02}. The problem is now addressed
to the decay dynamics in the $m$-gons and is discussed through the orbit
survival probability $\Psi _{m}(t)=|d(N_{tm}/N_{0m})/dt|$ related to the
number of the survived orbits $N_{tm}$. The decay spectra given by the $%
\varphi $-set regular-orbit numbers $N_{m}(t,\varphi )$, are also studied
for the case of small number of vertices $m$.

A general approach to the decay problem based on a simple decay kinetic
equation\cite{KN00} naturally arrives at the primary slow relaxation of the
regular-orbit sets given by $\Psi _{m}^{(\alpha )}\varpropto t^{-2}$ (see
Eq.(\ref{Nt})). We have demonstrated that the universal $\alpha $-channel,
attributed for both the chaotic\cite{KN00} and nonchaotic\cite{VK01}
billiards, is also characteristic of nonintegrable rational polygons. The
primary relaxation-motion regime originated from the piece-line parts of the
polygonal table is associated with long-living sliding orbits with large
collision angles $\varphi $. In the corresponding phase space these orbits
are unbounded trajectories (see {\bf Fig.4}) that move without splitting at
angle vertices along Bleher's principal corridors. Following to the simplest
polygonal orbit classification by Gutkin\cite{G96}, the regular
sliding-orbit sets can be presented by the
''infinite-past-to-infinite-future'' trajectories. They ''never'' hit
vertices, preserve the initial linear momenta, and show a regular behavior
in the orbit decay spectra $N_{m}(t,\varphi )$ (see Fig.1). Conversely, the
singular orbit sets caused by the ''infinite-past-to-vertex'', the
''vertex-to-infinite-future'', and the diagonal ''vertex-to-vertex''
trajectories\cite{trapped} exhibit pronounced weights $C_{m}(\varphi )$ in
the orbit-decay process (shown by high peaks in Fig.1). Eventually, they do
not play any significant role in the wall-collision statistics in $m$-gons
with small number of vertices limited by, approximately, $3\leq m\leq 8$,
and thus the primary relaxation dominates. This corroborates by our
numerical study (analyzed in Table 1and Fig.1) and, in general, is in accord
with studies of the closed polygons by the orbit-wall collision statistics%
\cite{KV02} and by the orbit complexity\cite{M00}.

When the number of vertices is large, the secondary relaxation with the
survival probability $\Psi _{m}^{(\beta )}\varpropto t^{-\beta -1}$
predominates over the primary relaxation (see Figs. 2,3). The established
domain for the decay exponent $1<\beta <2$ corresponds to that known\cite
{FS95} for the chaotic SB. Qualitatively, the survival probability function $%
\Psi _{m}^{(\beta )}$ can be associated with the distribution function for
trajectories trapped by the strange attractors, discussed in the theory of
the open classical chaotic systems, or with the corresponding waiting-time
probability function\cite{ZKS99}. With accounting of findings for the decay
dynamics on the SB by Fendrik's group\cite{FRS94,FS95}, one can expect that
the secondary relaxation is due to the {\em singular trapped} orbit sets
that move freely along Bleher's ''hidden corridors''. Similarly to the case
of the chaotic billiards, the observation conditions for the secondary
relaxation in rational polygons are sensitive to the initial conditions and
to the geometrical constrains. Indeed, the $\beta $-relaxation channel turns
up to be closed if the initial particle distribution is simulated\cite{Diss}
in the 2D {\em collision} subspace $\Omega _{cm}$. In the case of the 3D $%
\Omega _{m}$ space the secondary relaxation appears to be dynamically stable
under the geometrical constraint $m>m_{\alpha }$, where $m_{\alpha }=2\pi
R/\Delta $ is given by the $\alpha $-to-$\beta $-relaxation criterium
estimated in Eq.(\ref{m-alfa}). As shown, this criterium meets the favorable
survival conditions for the regular-motion regime with those induced by
rationality of the vortices. The latter are generated by the sliding orbits
through the vertex-''ordering'' effects and are associated with the
vortexlike orbits\cite{KV02}. As follows from Eq.(\ref{m-alfa}), the
observation window of such a motion disappears in the limit $\Delta
\rightarrow 0$, when the vortexlike orbits do not survive in the closed
polygons (see Fig.2 in Ref.\cite{KV02}). Finally, we have demonstrated that
the vortexlike orbits become stable in the open rational polygons and
visible through the secondary slow relaxation common for the chaotic systems.

ACKNOWLEDGMENTS

The authors are grateful to Josef Klafter for drawing their interest to the
escape problem in chaotic systems\cite{ZKS99}. Special thanks are due to
Mario Jorge Dias Carneiro for illuminating discussions. The financial
support of the Brazilian agency CNPq is also acknowledged.

\section{APPENDIX. ORBIT-SET COLLISION TIMES}

In a given $m$-gon a number of geometrically equivalent walls $k$ is bounded
above by

\begin{equation}
q_{m}=\left\{ 
\begin{array}{cc}
m & \text{for odd }m; \\ 
m/2 & \text{for even }m.
\end{array}
\right.  \eqnum{A1}
\end{equation}
The current collision angle $\varphi _{km}$ with a wall $k$ $(=1,2...q_{m})$
of a $\varphi $-orbit with $\varphi =[0,\pi /2q_{m}]$ is reduced through the
relation $\varphi _{km}=\varphi -\Theta _{km}$ with the help of $\Theta
_{km}=[-\pi /2,\pi /2]$ introduced as the lowest angle between the $k$-wall
and axis $x$ , namely 
\begin{equation}
\Theta _{km}=\frac{\pi }{2q_{m}}\left\{ 
\begin{array}{cc}
q_{m}-2k+1 & \text{for odd }q_{m}; \\ 
q_{m}-2k & \text{for even }q_{m}.
\end{array}
\right.  \eqnum{A2}
\end{equation}
As shown in Fig.4 for the particular case $m=3$, the estimates for the
wall-collision times $t_{cm}(\varphi )$ are found through summation of
numbers of intersections $n(t,\varphi _{km})$ for a trajectory, induced by a
given $\varphi $-set orbit, considered in the correspondent infinite LG
lattice, namely 
\begin{equation}
n_{cm}(t,\varphi )\equiv \frac{t}{t_{cm}(\varphi )}=\sum%
\limits_{k=1}^{q_{m}}n(t,\varphi _{km})  \eqnum{A3}
\end{equation}
The estimation procedure can be exemplified by a relation $t\cos (\varphi
_{13})=n(t,\varphi _{13})3a_{3}$. The latter employes the fact that a
distance between the equivalent walls is $3a_{3}$, where $a_{m}=R\cos (\pi
/m)$ stands for the {\em apothem} in a given $3$-gon. This yields

\begin{equation}
t_{cm}(\varphi )=a_{m}q_{m}\left[ \sum\limits_{k=1}^{q_{m}}\cos (\varphi
-\Theta _{km})\right] ^{-1}  \eqnum{A4}
\end{equation}
where $q_{m}$ and $\Theta _{km}$ are given in Eqs.(A1) and (A2),
respectively. Straightforward estimation of Eq.(A4)\cite{estimation} results
in the collision times $t_{cm}(\varphi )$ given in Eq.(\ref{tc-m}).

{\large FIGURE CAPTIONS}

.

Fig. 1. Analysis of the algebraic decay simulated in pentagon ($5$-gon) and
heptagon ($7$-gon) within the collision $\Omega _{cm}$ space. {\em Symbols }%
- numerical data on the partial weights $C_{5}^{(\exp )}(\varphi )$ and $%
C_{7}^{(\exp )}(\varphi )$ deduced from the observed spectra of the survived
orbits $N_{t5}^{(\exp )}(\varphi )$ and $N_{t7}^{(\exp )}(\varphi )$ with
the help of Eq.(\ref{Nt}) and simulated for $\varphi $-set orbits with $%
0\leq \varphi \leq \varphi _{m}$ at distinct times $t=20,30\tau _{em}$ . 
{\em Line: }the overall-collision-anlle weight $C_{cm}^{(\exp )}$.

{\em Insert left:} {\em Points }- data $N_{t5}$ for the overall survived
orbits at late times and their regular-orbit analysis with the help of Eq.(%
\ref{Nt}).

{\em Insert right:} {\em Points }- data for $\varphi $-set collision time $%
t_{c5}(\varphi )$ simulated within the reduced domain $0\leq \varphi \leq
\pi /10$.{\em \ Line }- the same predicted in Eq.(\ref{tc-m}).

.

Fig. 2. Temporal evolution of the survived orbits in rational polygons with
small opening width ($\Delta =0.05R$) against the reduced time in log-log
coordinates. Reduction is given by the help of Eq.(\ref{tau/ec}) for the
escape time $\tau _{e}=300$, chosen common for all cases. {\em Points}:
numerical data for decay of the $\Omega _{m}$ space phase simulated by $%
N_{0}=10^{6}$ particles in the $m$-gons (squares) and the correspondent CB
(circles).

.

Fig. 3. Temporal evolution of the survived orbits in rational polygons with
large opening width ($\Delta =0.20R$) against the reduced time in log-log
coordinates. Notations are the same as in Fig.2.

.

Fig.4. Estimation of the $\varphi $-orbit collision time $t_{cm}(\varphi )$
on the bases of Eq.(A3) for the case of $m=3$. The regular piece-line orbit $%
a,b,c,d,e...$ is represented by the infinite straight-line trajectory in the
triangle LG lattice with the intersection-point sequences $%
1,2,3,4...n(t,\varphi _{km})$. The equivalent walls $k$, the unreduced
collision angles $\varphi _{km}$, and the axillar angles $\Theta _{km}$ are
shown.\newpage

.

{\large TABLE 1}

.

\begin{tabular}{||c||c|c|c|}
\hline\hline
$m$ & $C_{cm}^{(\text{tot})}$ & $C_{cm}^{(\text{reg})}$ & $D_{cm}^{(\text{tot%
})}$ \\ \hline\hline
3 & 0.140 & 0.116 & 0.135 \\ \hline
4 & 0.220 & 0.219 & 0.210 \\ \hline
5 & 0.094 & 0.086 & 0.090 \\ \hline
6 & 0.149 & 0.139 & 0.150 \\ \hline
7 & 0.092 & 0.069 & 0.090 \\ \hline
8 & 0.099 & 0.096 & 0.100 \\ \hline
\end{tabular}

.

Table 1. Fitting parameters of the temporal algebraic decay of the collision
space $\Omega _{cm}$ simulated in the weakly open $m$-gons with $\Delta
=0.05R$. {\em Notations:} $C_{cm}^{(\text{tot})}$ and $C_{cm}^{(\text{reg})}$
correspond to the data on $\varphi $-sets observed in the decay spectra $%
C_{cm}^{(\text{exp})}(\varphi )$ (see Fig. 1) and averaged over,
respectively, all collision angles and with excluding singular-orbit angles
manifested by the high peaks; $D_{cm}^{(\text{tot})}=D_{cm}^{(\text{exp})}$-
the overall-set weights of the algebraic tail given in Eq.(\ref{Nt}) and
derived within the primary relaxation window (see the left insert in Fig.1).

.


\begin{references}
\bibitem{FKC01}  An exception was recently claimed by N. Friedman et al. in
Phys. Rev. Lett. {\bf 86}, 1518 (2001) for the case of the ''real''
atom-optics ''titled''-stadium-shaped billiard, where only the short-time
exponential decay was observed. No explanations were given why the late-time
decay regime, which is controlled by the boundary hole, was not achieved.
\bibitem{DynEqui}  With accounting that the vertex angles are given by $\Phi
_{m}=\pi (1-2/m)$ in a $m$-gon, one can see that any orbit is presented by $m
$ (or by $m/2$) distinct collision angles for even (or odd) $m$ and, hence,
all $m$ (or $m/2$) sides are dynamically equivalent. See also footnote $^{3)}
$ in Ref.\cite{VK01}.
\bibitem{exept}  Exception should be given for the case of $m=4$ discussed
in details in Ref.\cite{VK01}.
\bibitem{trapped}  Note that the diagonal ''vertex-to-vertex'' singular
trajectories may remain trapped in the even-$m$-gons. The ''bouncing-ball''
trajectories, formed by ''infinite-past-to-vertex '' and by
''vertex-to-infinite-future'' trajectories, can decay through both the $%
\alpha $ and $\beta $ channels.
\bibitem{estimation}  A geometric correspondence between a given $m$-fold
rotational- symmetry polygon and the infinite-range periodically translated
LG lattice exists only in the cases $m=3,4$ and $6$. Meanwhile, one can show
that Eq.(A4) works well for the long-living orbits, when $t\gg a_{m}q_{m}$.
This follows from the same estimation procedure shown in Fig.4 and is
supported by our numerical analysis illustrated in the right insert in
Fig.1.\newpage
\end{references}
\end{document}